\documentclass[aps,preprint,letterpaper,superscriptaddress,longbibliography]{revtex4-2}
\usepackage[justification=raggedright,singlelinecheck=false,labelfont=bf,labelsep=period]{caption,subcaption}
\usepackage{braket}
\usepackage{siunitx}
\usepackage{comment}
\usepackage{nameref}
\usepackage[version=4]{mhchem}

\usepackage{scalerel}
\usepackage{tikz}
\usetikzlibrary{svg.path}
\definecolor{orcidlogocol}{HTML}{A6CE39}
\tikzset{
  orcidlogo/.pic={
    \fill[orcidlogocol] svg{M256,128c0,70.7-57.3,128-128,128C57.3,256,0,198.7,0,128C0,57.3,57.3,0,128,0C198.7,0,256,57.3,256,128z};
    \fill[white] svg{M86.3,186.2H70.9V79.1h15.4v48.4V186.2z}
                 svg{M108.9,79.1h41.6c39.6,0,57,28.3,57,53.6c0,27.5-21.5,53.6-56.8,53.6h-41.8V79.1z M124.3,172.4h24.5c34.9,0,42.9-26.5,42.9-39.7c0-21.5-13.7-39.7-43.7-39.7h-23.7V172.4z}
                 svg{M88.7,56.8c0,5.5-4.5,10.1-10.1,10.1c-5.6,0-10.1-4.6-10.1-10.1c0-5.6,4.5-10.1,10.1-10.1C84.2,46.7,88.7,51.3,88.7,56.8z};}}
\newcommand\orcidicon[1]{\href{https://orcid.org/#1}{\mbox{\scalerel*{
\begin{tikzpicture}[yscale=-1,transform shape]
\pic{orcidlogo};
\end{tikzpicture}
}{|}}}}

\usepackage{etoolbox}

\makeatletter
\pretocmd\frontmatter@thefootnote{\color{blue}}{}{}

\usepackage{dcolumn} 
\usepackage{titlesec}
\titlespacing*{\section}
{0pt}{4ex}{1.3ex}
\titlespacing*{\section}
{0pt}{4ex}{0ex}

\titleformat{\section}
  {\bfseries\MakeUppercase}{\thesection\raggedright}{1em}{}
  \titleformat{\section}
 {\bfseries}{\thesection\raggedright}{1em}{}

\usepackage{xstring}
\usepackage{xspace}

\usepackage[utf8]{inputenc}
\usepackage[english]{babel}
\usepackage{gensymb}
\usepackage{mathtools}

\usepackage{float}
\usepackage{graphicx}

\usepackage{url}
\usepackage[utf8]{inputenc}
\usepackage[T1]{fontenc}
\usepackage{textcomp}
\usepackage{gensymb}

\usepackage[colorlinks,citecolor=red,urlcolor=blue,bookmarks=false,hypertexnames=true]{hyperref}
\usepackage{cleveref}

\begin{document}

\title{Isotropic atomic layer etching of MgO-doped lithium niobate using sequential exposures of H$_2$ and SF$_6$/Ar plasmas}

\author{Ivy I. Chen~\orcidicon{0009-0009-0942-5658}}

\affiliation{%
Division of Engineering and Applied Science, California Institute of Technology, Pasadena, California 91125, USA
}%
\author{Jennifer Solgaard}
\affiliation{%
Department of Electrical Engineering, California Institute of Technology, Pasadena, California 91125, USA
}%
\author{Ryoto Sekine}
\affiliation{%
Department of Electrical Engineering, California Institute of Technology, Pasadena, California 91125, USA
}%
\author{Azmain A. Hossain}
\affiliation{%
Division of Engineering and Applied Science, California Institute of Technology, Pasadena, California 91125, USA
}%
\author{Anthony Ardizzi}
\affiliation{%
Division of Engineering and Applied Science, California Institute of Technology, Pasadena, California 91125, USA
}%
\author{David S. Catherall}
\affiliation{%
Division of Engineering and Applied Science, California Institute of Technology, Pasadena, California 91125, USA
}%

\author{Alireza Marandi}
\affiliation{%
Department of Electrical Engineering, California Institute of Technology, Pasadena, California 91125, USA
}%
\author{James R. Renzas}
\affiliation{%
Oxford Instruments Plasma Technology, North End, Bristol BS49 4AP, United Kingdom
}%
\author{Frank Greer}
\affiliation{%
Jet Propulsion Laboratory, California Institute of Technology, Pasadena, California 91109, USA
}%

\author{Austin J. Minnich~\orcidicon{0000-0002-9671-9540} }
 \email{aminnich@caltech.edu}
\affiliation{%
Division of Engineering and Applied Science, California Institute of Technology, Pasadena, California 91125, USA
}%

\date{\today}
               
\begin{abstract}

Lithium niobate (LiNbO$_3$, LN) is a ferroelectric crystal of interest for integrated photonics owing to its large second-order optical nonlinearity and the ability to impart periodic poling via an external electric field.  However, on-chip device performance based on thin-film lithium niobate (TFLN) is presently limited by propagation losses arising from surface roughness and corrugations. Atomic layer etching (ALE) could potentially smooth these features and thereby increase photonic performance, but no ALE process has been reported for LN. Here, we report an isotropic ALE process for $x$-cut MgO-doped LN using sequential exposures of H$_2$ and SF$_6$/Ar plasmas. We observe an etch rate of $1.59 \pm 0.02$ nm/cycle with a synergy of $96.9$\%. We also demonstrate ALE can be achieved with SF$_6$/O$_2$ or Cl$_2$/BCl$_3$ plasma exposures in place of the SF$_6$/Ar plasma step with synergies of $99.5$\% and $91.5$\% respectively. The process is found to decrease the sidewall surface roughness of TFLN waveguides etched by physical Ar$^+$ milling by 30\% without additional wet processing. Our ALE process could be used to smooth sidewall surfaces of TFLN waveguides as a post-processing treatment, thereby increasing the performance of TFLN nanophotonic devices and enabling new integrated photonic device capabilities. 


\end{abstract}
\maketitle

\section{Introduction}

Lithium niobate (LiNbO$_3$ or LN) is a ferroelectric crystal of interest for a variety of integrated photonics applications ranging from electro-optic modulators in fiber-optic communications to quantum optics. \cite{Boes:unlocking} LN is a trigonal crystal characterized by a threefold rotational symmetry about the crystallographic $z$ axis. Because $x$-cut electro-optic modulators have fewer processing requirements compared to their $z$-cut counterparts, \cite{Zhang:21:xzcut} the $x$-cut surface is the relevant surface for LN nanophotonic circuits. The crystal structure of LN is described in Refs.~\cite{iyi_LN_defect_1995, sanna_linbo3_2017, sanna_lithium_2010}. LN exhibits a number of desirable properties for photonics, including a large transparency window, wide electro-optic bandwidth, ferroelectric properties, and high second-order nonlinear susceptibility, \cite{Weis1985_LN_properties, wong2002_LN_properties, kong2020LN, volk2008LN, arizmendi2004LNphotonic} making it an attractive platform compared to other materials like silicon nitride. \cite{Zhu:21} By incorporating > 5\% molar concentration MgO into the melt during the Czochralski crystal growth process, the optical damage threshold is raised, allowing for high-intensity photonic applications \cite{bryan_optical_damage_MgO_1984}. 


 Early efforts to create on-chip photonic devices involved Ti ion diffusion or proton-exchange on bulk LN wafers to provide the necessary refractive index contrast. \cite{Hu_PE_LN_etch,PE_LN_Cabrera_HLiCorr,cabrera_hydrogen_ln, Dorrer_PE_LN_diffusion, Cai_PE_LN_EO_properties, Ren_PE_LN_Plasma} However, the relatively small refractive index contrast from this approach resulted in weak optical confinement, imposing limitations on the types of devices and nonlinear phenomena that could be observed. With the development of ion-slicing and wafer bonding processes for LN on silicon dioxide, \cite{LN_ion_slicing, Levy_Osgood_2000_patent, LN_wafer_bond} thin-film lithium niobate (TFLN) wafers have become commercially available, allowing for the realization of dense circuits with tightly-confining waveguides. Devices that have been fabricated on TFLN include squeezed quantum states on-chip, \cite{Marandi_2022_squeezed_states} $>$100 GHz electro-optic modulators with CMOS compatible voltages, \cite{Wang2018_LN_modulator_CMOS}  broadband frequency comb sources, \cite{sekine2023multioctave, wu2023_broadband_fc,Jankowski_broadband_optcis} and on-chip ultra-fast lasers. \cite{Yu2022_modelocked, guo2023modelocked} 
 

 A necessary step in LN device fabrication is pattern transfer, typically using a dry etching process. 
 Process development for dry etching of LN is more challenging compared to that for other photonic materials such as SiN because LN is a ternary compound. Fluorine- \cite{F_RIE_LN_proton_exchange_Hu} or chlorine- \cite{Cl_RIE_LN} based reactive ion etching (RIE) processes have been reported, but they suffer from redeposition of non-volatile Li compounds such as LiF, leading to an increase in sidewall roughness and scattering loss, which is the dominant loss mechanism. \cite{Zhu:21,F_RIE_LN_proton_exchange_Hu} Proton-exchanged LN has been noted to have lower LiF redeposition during plasma etching due to lower surface Li content. Deep (> 1 $\mu$m) fluorine-based etches with less LiF redeposition have been accomplished with proton-exchanged LN. \cite{Hu_PE_LN_etch, Jun_2012_PE_LN_Etch, Ren_2008_PE_LN_Etch, Aryal_LN_Proton}

In the device community, physical Ar$^+$ milling remains the preferred dry etch method used for pattern transfer. However, this method has its own limitations such as low etch selectivity with common lithography resists, non-vertical sidewalls, redeposition of LN, and variations of etch depth across a single chip. \cite{Zhu:21, Ar_LN_Etch_Clean} Various approaches are available to remedy some of these limitations; for instance, redeposited LN after Ar$^+$ milling is typically removed using an RCA clean. However, the wet process also introduces corrugations in periodically-poled LN (PPLN) due to differential wet etch rates between poled domains,  \cite{Wang:18_PPLN} leading to optical loss which dominates the overall loss in TFLN devices. \cite{Kippenberg_Loncar_MaterialQ} As a result, various device figures of merit such as resonator quality factors are at least an order of magnitude from their intrinsic upper limits. Decreasing losses associated with corrugations and sidewall roughness in PPLN circuits will enable system-level integration of on-chip nonlinear optics and allow for quantum information processing. \cite{Zhu:21}

These challenges could be addressed with improved nanofabrication techniques which offer sub-nanometer-scale etch depth control and surface smoothing. In particular, thermal or plasma-enhanced atomic layer etching (ALE) has demonstrated etch depth control on the angstrom scale and an ability to smooth surfaces to the sub-nanometer scale. \cite{kanarik_rethinking_etch, george_thermal_ALE} ALE consists of sequential, self-limiting surface chemical processes that lead to etch per cycles ranging from fractional monolayers to a few monolayers in crystalline materials. ALE can be anisotropic (directional) or isotropic (thermal or plasma-thermal). \cite{george_thermal_ALE, kanarik_ALE_overview, Sang_2020} Anisotropic ALE is based on surface modification by adsorption of a reactant followed by low-energy ion or neutral atom sputtering. \cite{Horiike_Digital_etch, kanarik_ALE_overview, Oehrlein_tipping_pt_2015} The self-limiting nature of anisotropic ALE is defined by the thickness of the modified surface and the difference in sputtering threshold between the modified and unmodified surface. Thermal (isotropic) ALE is based on a cycle of surface modification and volatilization reactions. Recent developments in ALE have also employed a pulsed-bias approach, where the flow of gases is held constant and the DC bias is turned on and off, resulting in faster ALE cycle times. \cite{Julian_pulsed_bias} Thermal and anisotropic ALE recipes have been developed for various semiconductors and dielectrics such as SiO$_2$, \cite{dumont_ALE_silica_HF_TMA, rahman_ALE_silica_TMA_CHF3} InP, \cite{Ko_InP_ALE, Park_InP_ALE, Park_InP_ALE_2} GaAs, \cite{lu_ALE_InGaAs,MEGURO1994_GaAs_ALE,Aoyagi_1992_GaAs_ALE,Aziziyan2019_GaAs_ALE} and Si$_3$N$_4$. \cite{Li_SiN_ALE,Thermal_SiN_ALE,Ishii_2017_SiN_ALE,Sherpa_2016_SiN_ALE,Matsuura_1999_SiN_ALE, DanShanks_QALE_SiN}
Surface smoothing due to ALE has been observed for various materials, \cite{kanarik_rethinking_etch, Zywotko_RALE_alumina_HF_TMA_nopurge, ohba_ALE_GaN_AlGaN, kanarik_ALE_synergy, 2023_Azmain_TiN, Haozhe_2023_AlN, DanShanks_QALE_SiN, konh_selectivity_2022} a feature which has been attributed to conformal layer-by-layer removal and curvature-dependent surface modification. \cite{gerritsen_ALD_ALE_surface_smoothing} Despite the potential to smooth step pattern corrugations and sidewall roughness in PPLN, no ALE processes have been reported for LN. 

Here, we report an isotropic ALE process for MgO-doped $x$-cut LN. Using sequential exposures of H$_2$ and SF$_6$/Ar plasmas, we measure an etch per cycle (EPC) of $1.59 \pm 0.02$ \AA/cycle with a synergy of $96.9\%$. We observe the saturation of both half-steps of the process. While surface roughness is observed to increase on flat surfaces, a 30\% reduction in surface roughness on waveguide sidewalls is observed after 50 cycles of ALE. In addition, we demonstrate that the SF$_6$/Ar plasma step can be replaced with an O$_2$/SF$_6$ or Cl$_2$/BCl$_3$ plasma and achieve EPCs of $2.24 \pm 0.02$ and $1.65 \pm 0.03$ and synergies of 99.5\% and 91.5\%, respectively. The process could be used as a post-processing step after Ar$^+$ milling to smoothen sidewall roughness and corrugations in periodically-poled TFLN devices and thereby enhance their photonic performance.

\section{Experimental Methods}

\begin{figure}
    \centering
    {\includegraphics[width=\textwidth]{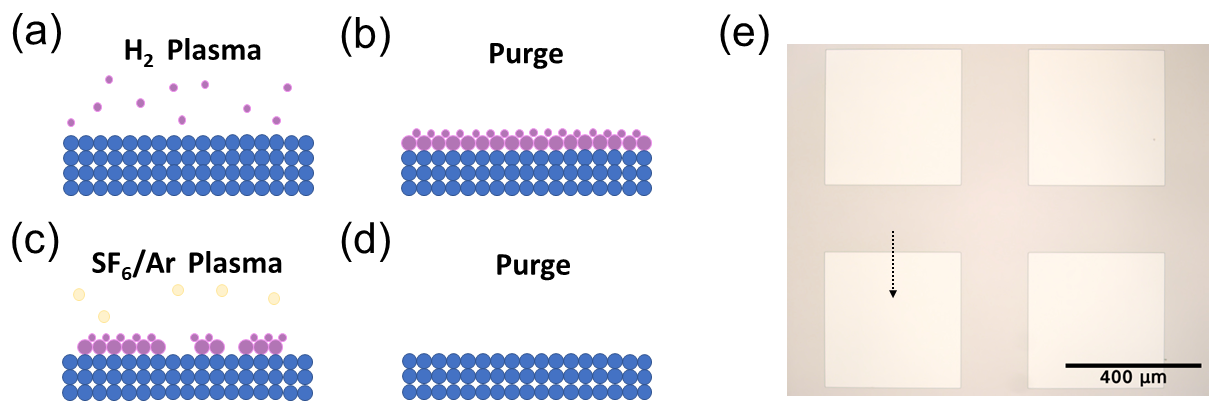}
    \phantomsubcaption\label{fig:ale_recipe_a}
    \phantomsubcaption\label{fig:ale_recipe_b}
    \phantomsubcaption\label{fig:ale_recipe_c}
    \phantomsubcaption\label{fig:ale_recipe_d}
    \phantomsubcaption\label{fig:litho}
    }
    \caption{(a-d) ALE process for MgO-doped LN. A hydrogen plasma exposure (pink) leads to a hydrogen-rich modified layer (pink circles) at the top of the sample (blue dots), followed by a  purge. A subsequent SF\textsubscript{6}/Ar plasma exposure (yellow dots) yields volatile species. A final purge completes the cycle. (e) Microscope image ($10 \times$ magnification) of the developed lithography pattern on the LN wafer. The dotted line indicates the direction of AFM scan for etch depth measurements.}
    \label{fig:ALE_recipe}
\end{figure}

The samples consisted of bulk 3-inch 5\% mol MgO-doped LN wafers (G \& H Photonics). The wafers were diced into 7 mm $\times$ 7 mm substrates using a Disco DAD 321 dicing saw and then cleaned by  sonication in AZ NMP Rinse, acetone, and isopropyl alcohol. The samples were etched in an Oxford Instruments PlasmaPro 100 Cobra system configured for ALE. As shown in \Cref{fig:ale_recipe_a,fig:ale_recipe_b,fig:ale_recipe_c,fig:ale_recipe_d}, the process consisted of sequential exposures to H$_2$ and SF$_6$/Ar plasmas with purges between each exposure. This process was motivated by the observation that proton-exchanged LN can be etched with fluorine plasmas with reduced LiF redeposition \cite{Ren_PE_LN_Plasma,Hu_PE_LN_etch,Aryal_LN_Proton, Jun_2012_PE_LN_Etch, Ren_2008_PE_LN_Etch} and because the same plasmas successfully achieved quasi-ALE of SiN. \cite{DanShanks_QALE_SiN}

The nominal ALE recipe consists of an 40-second H$_2$ plasma exposure (300 W ICP power, 52.5 W RIE power, 209 V DC bias, 50 sccm H$_2$) followed by a 40-second SF$_6$/Ar exposure (300 W ICP power, 3.5 W RIE power, 50 V DC bias, 17 sccm SF$_6$, 35 sccm Ar). The effect of EPC on RF bias power was not studied. 5 second purge times with 40 sccm Ar were used between plasma half-steps. The chamber pressure was set at 10 mTorr and the substrate table was cooled to \SI{0}{\degreeCelsius} using liquid nitrogen as measured by the table thermometer.

To measure saturation curves, the chamber pressure and ICP power were kept constant at 10 mTorr and 300 W respectively, while the exposure time for each half-step was varied. H$_2$ plasma exposure time was varied from 0 to 50 seconds with SF$_6$/Ar plasma held at 30 seconds, and SF$_6$/Ar plasma exposure time was varied from 0 to 50 seconds with H$_2$ plasma exposure held at 30 seconds. Prior to introducing the sample into the chamber for etching, the etching chamber was cleaned with a blank Si wafer and a 30-minute Ar$^+$ plasma with 1500 W ICP and 100 W RF power followed by a 15-minute O$_2$/SF$_6$ plasma with the same power parameters. After the sample was loaded into the chamber, a 3-minute wait time was used before processing to allow the sample to thermally equilibrate with the table. All samples were etched for 50 cycles unless otherwise noted. After etching, the photoresist was removed using room temperature AZ NMP Rinse for at least 30 minutes to ensure complete removal of the resist, followed by sonication in acetone and isopropyl alcohol.

To enable etch depth measurements, step patterns consisting of periodic $400 \times 400$ $\mu m^2$ squares were written into a resist using photolithography, as shown in \Cref{fig:litho}. The pattern was transferred to the +$x$ face of the samples using AZ5214 photoresist and a Heidelberg MLA 150 Maskless Aligner with a dose of 150 mJ/cm$^2$, followed by development with AZ 300 MIF developer. Etch per cycle (EPC) was calculated by measuring the difference in height from etch depth for a processed sample and dividing it by the total number of cycles. AFM scans were performed on a Bruker Dimension Icon atomic force microscope (AFM) to measure total etch depth and surface roughness. The total etch depth was measured using $2.5 \times 10$ $\mu$m$^2$ AFM scan with the scan rate set to 0.5 Hz. The step profile was averaged over the entire scan using Nanoscope Analysis 1.9 software to obtain the etch depth. RMS surface roughness of a reference TFLN Ar$^+$ milled waveguide sample and power spectral density (PSD) scans were obtained over a 50 $\times$ 50 nm$^2$ area with a 0.5 Hz scan rate. Waveguide sidewall slope on measured TFLN samples and sample tilt from all AFM scans were removed via quadratic plane fit.

X-ray photoelectron spectroscopy (XPS) analysis was performed using a Kratos Axis Ultra x-ray photoelectron spectrometer using a monochromatic Al K$\alpha$ source. A 1.69 nm thick layer of carbon, as measured by a quartz crystal monitor, was deposited using sputtering to reduce charging effects during scans (Leica EM ACE600 Carbon Evaporator). The resulting data was analyzed in CASA-XPS from Casa Software Ltd. For each sample, we collected the carbon C1s, oxygen O1s, niobium Nb3d$_{5/2}$ and Nb3d$_{3/2}$, niobium Nb4s, lithium Li1s, fluorine F1s, and magnesium Mg2p peaks. The carbon C1s peak was used as a reference to calibrate peak positions. We fit the data using a Shirley background subtraction and peak fitting routines from Refs.~\cite{2002_LN_H2_XPS, gruenke2023LNXPS}. 

Two alternate recipes were also investigated. The first alternate recipe consists of a 40-second H$_2$ plasma exposure of the same parameters as the SF$_6$/Ar recipe followed by a 40-second O$_2$/SF$_6$ exposure (300 W ICP power, 3.5 W RIE power, 39 V DC bias, 35 sccm O$_2$, 15 sccm SF$_6$). The second alternate recipe uses the same 40-second H$_2$ plasma exposure followed by a 40-second Cl$_2$/BCl$_3$ exposure (300 W ICP power, 5 W RIE power, 73 V DC bias, 20 sccm Cl$_2$, 40 sccm BCl$_3$). The second alternate recipe was motivated by reports of ALE processes for metal oxides based on BCl\textsubscript{3}, \cite{Oehrlein_tipping_pt_2015} and the Cl$_2$:BCl$_3$ gas flow ratio was selected based on an RIE recipe of LN using chlorine.  \cite{Cl_RIE_LN}  Whether the O$_2$/SF$_6$ and Cl$_2$/BCl$_3$ processes were at saturation was not determined.  Etch depth measurements and 500 $\times$ 500 nm$^2$ surface roughness scans over 20 cycles from these alternate processes were compared with 20 cycles of the original ALE recipe consisting of a 40-second H$_2$ plasma exposure followed by a 40-second SF$_6$/Ar exposure.

\section{Results}

\begin{figure}
    \centering
    
{\includegraphics[width = \textwidth]{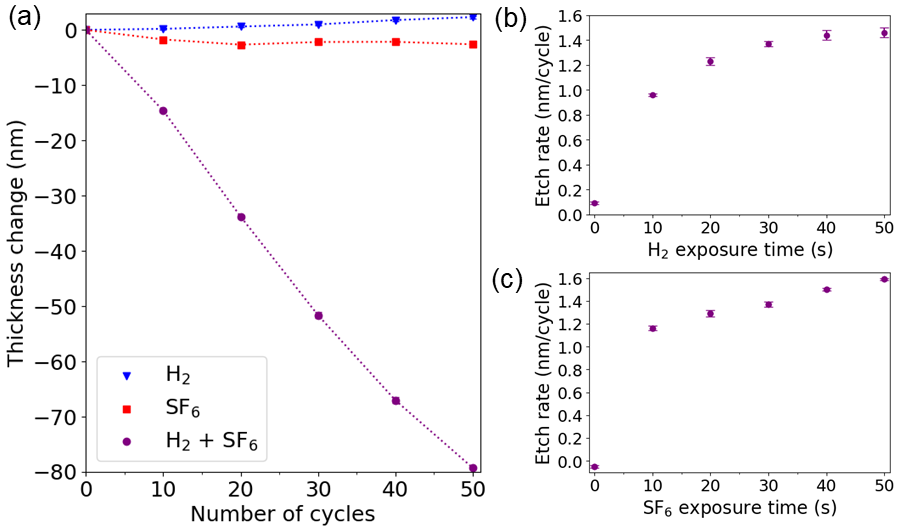}
    \phantomsubcaption\label{fig:EPC_cyclenum}
    \phantomsubcaption\label{fig:H2_selflimit}
    \phantomsubcaption\label{fig:SF6_selflimit}
}
\caption{(a) EPC versus cycle number with 40 second H$_2$ plasma exposures only (blue triangles), 40 second SF$_6$/Ar plasma exposures only (red squares), and both half-cycles (purple circles). All processes occur at \SI{0}{\degreeCelsius}. The dashed lines are guides to the eye. (b) EPC versus H$_2$ plasma exposure time with SF$_6$/Ar plasma exposure time fixed at 30 s. (c) EPC versus SF$_6$/Ar plasma exposure time with H$_2$ plasma
exposure time fixed at 30 s. The etch rates are observed to saturate with exposure time, demonstrating the
self-limiting nature of the process.
}
\label{fig:self-limiting} 
\end{figure}

\Cref{fig:EPC_cyclenum} shows the thickness change of LN versus cycle number for individual half cycles and the overall process. An etch rate of 0.06 nm/cycle is observed for the SF$_6$/Ar plasma half step. For the H$_2$ plasma step, a thickness increase was observed, which might be attributed to a volume expansion due to amorphization of the crystal during the H$_2$ plasma exposure. Such thickness increases for one half-step have been reported in other processes. \cite{george_thermal_ALE} On the other hand, when using both steps sequentially, an etch rate of 1.59$\pm$0.02 nm/cycle is observed. 


To gain more insight into the process and verify its self-limiting nature, we measured saturation curves for each half-cycle. In \Cref{fig:H2_selflimit}, the SF$_6$/Ar plasma half step is held constant at 30 seconds while the H$_2$ plasma exposure time is varied from 0 to 50 seconds. Saturation occurs at $1.46 \pm 0.04$ nm/cycle above 30 seconds H$_2$ plasma exposure time. In \Cref{fig:SF6_selflimit}, the H$_2$ plasma exposure time is held constant at 30 seconds while the SF$_6$/Ar plasma exposure time is varied from 0 to 50 seconds. The etch rate exhibits a soft saturation, as the etch rate continues to increase with increasing exposure time. For SF$_6$/Ar exposure times longer than 30 seconds, the etch rate continues to increase at a rate of 0.1 nm/cycle per 10 seconds of additional SF$_6$/Ar plasma exposure, indicating that the half step exhibits soft saturation. Soft saturation with SF$_6$/Ar plasma has been reported previously and was attributed to the diffusion-limited fluorination of the surface  \cite{chittock_ALE_GaN_softsat}. In the present case, soft saturation is hypothesized to occur due to the presence of a concentration gradient of hydrogen into the LN film after H$_2$ plasma exposure. By increasing the SF$_6$/Ar plasma exposure time, more of the hydrogenated surface is removed, resulting in a soft-saturating curve. At 50 seconds SF$_6$ plasma exposure time, the etch rate is $ 1.59 \pm 0.02$ nm/cycle. The observation of saturation for both half-steps indicates that the process is indeed atomic layer etching.

The synergy, $S$, as defined by Ref.~\cite{kanarik_ALE_synergy}, quantitatively compares the etch depth using only individual steps of the ALE cycle to the etching obtained with the full etch cycle as $S=(1 - (\alpha+\beta)/EPC)\times 100$, where $\alpha$ and $\beta$ are the etch rate of the H$_2$ plasma and  SF$_6$/Ar half-cycles, respectively; and $EPC$ is the etch rate of the full cycle. For the present process in which a thickness increase is observed after H$_2$ plasma exposure, we take a conservative approach and calculated the synergy assuming zero EPC for that step. We obtain a synergy value of  96.9\% for the nominal recipe. This synergy value is comparable with typical synergy values reported in Ref.~\cite{kanarik_rethinking_etch}. 

We also investigated alternate ALE recipes using O$_2$/SF$_6$ or Cl$_2$/BCl$_3$ plasma exposures for the removal step. The O$_2$/SF$_6$ ALE process yielded an etch rate of $2.24 \pm 0.01$ nm/cycle over 20 cycles. The half-step EPCs for the H$_2$ and O$_2$/SF$_6$ plasma step are -0.04 and 0.01 nm/cycle, respectively. The synergy for this process is 99.5\%, with the H$_2$ plasma half step assumed to be 0 EPC for purposes of calculation as previously noted. The Cl$_2$/BCl$_3$ ALE process yielded an EPC of $1.65 \pm 0.03$  nm/cycle over 20 cycles; the half-step EPCs for the H$_2$ and Cl$_2$/BCl$_3$ plasma step are -0.04 and 0.14 nm/cycle, respectively, and the synergy for this process is 91.5\%. While the reaction mechanisms of the three processes were not studied in this work, the possible reactions are hypothesized to be fluorine or chlorine radicals forming  volatile compounds such as NbF$_5$, NbOF$_3$, OF$_2$,  NbOCl$_x$, and BOCl$_x$ as occurs in RIE of LN \cite{Cl_RIE_LN, 2021_LN_Fcompounds}.

\begin{figure}
    \centering
    
{\includegraphics[width = 450pt]{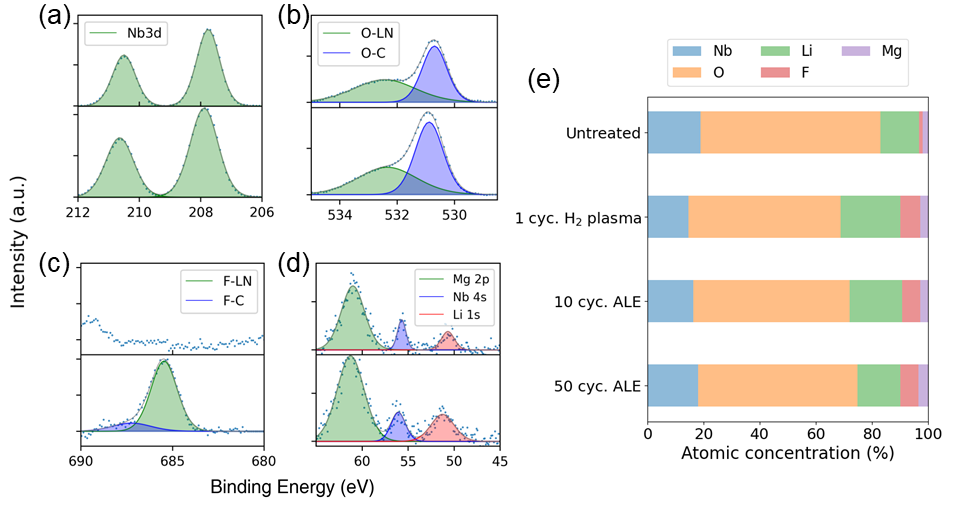}
    \phantomsubcaption\label{fig:xps_nb}
    \phantomsubcaption\label{fig:xps_o}
    \phantomsubcaption\label{fig:xps_f}
    \phantomsubcaption\label{fig:xps_mg}
    \phantomsubcaption\label{fig:xps_histogram}
}
\caption{Surface XPS spectra showing (a) Nb3d, (b) O1s, (c) F1s spectra, and (d) Nb4s, Li1s, Mg2p. The spectra is shown for (top) original and (bottom) etched bulk MgO-doped LN over 50 ALE cycles consisting of a 40 second H$_2$ plasma exposure followed by a 40 second SF$_6$/Ar exposure. The measured (blue dots) and fit spectra (gray lines) intensity are reported in arbitrary units (a.u.) against the binding energy on the x-axis. (e) Surface atomic concentration normalized by carbon atomic concentration from XPS spectra for each sample for untreated bulk LN, 1 cycle H$_2$ plasma exposure, 10 ALE cycles, and 50 ALE cycles.}
\label{fig:xps} 
\end{figure}

\begin{table}
\caption{\label{tab:xps_table} Atomic concentrations for the fitted XPS data.}
\begin{ruledtabular}
\begin{tabular}{cccccc}
 Sample & Nb (\%) & O (\%) & Li (\%) & F (\%) & Mg(\%) \\
\hline
Untreated   &  $18.92 \pm 0.22$    &    $ 64.17 \pm 0.76 $      &        $ 13.68 \pm 1.17 $    &  $1.32 \pm 0.56$  & $1.91 \pm 0.18$\\
1 half cycle H$_2$ plasma   &  $14.69 \pm 0.26$ &   $54.14 \pm 0.97$  &      $21.17 \pm 2.15$   &  $7.29 \pm 0.21$  &  $2.72 \pm 0.44$\\
10 cycles ALE  &  $16.40 \pm 0.22$ &  $55.64 \pm 0.76$   &     $18.79 \pm 1.62$     &  $6.38 \pm 0.19$ &  $2.79 \pm 0.32$ \\
50 cycles ALE  &  $18.04 \pm 0.17$  &  $56.83 \pm 0.55$  &     $15.17 \pm 1.15$  &  $6.56 \pm 0.13$ & $3.40 \pm 0.19$
\end{tabular}
\end{ruledtabular}
\end{table}

We next characterize the chemical composition of bulk LN before and after 50 cycles ALE for the SF$_6$/Ar plasma process using XPS. No depth-profiling XPS is reported due to preferential sputtering of O over Nb with an Ar$^+$ beam, \cite{Nb_pref_sput_XPS} complicating the interpretation of the measurements. The C1s peak at 284.8 eV is used as a reference. Binding energy values are reported in \Cref{tab:xps_table}. In \Cref{fig:xps_nb,fig:xps_o,fig:xps_f,fig:xps_mg}, we show the core levels of Nb3d, O1s; F1s; and Nb4s, Li1s, and Mg2p, respectively. For the Nb3d XPS spectra in \Cref{fig:xps_nb}, we observe a single doublet peak consisting of a 3d$_{5/2}$ and 3d$_{3/2}$ subpeak corresponding to LN (207.7 eV and 210.5 eV). \cite{LN-Nb3d_peaks_Kaufherr, database2000nist, 2016_XPS_LN_ratios} In \Cref{fig:xps_o}, we report the O1s spectra with two subpeaks at  530.7 and 532.4 eV, corresponding to metal oxide and O-C bonds, respectively. \cite{chastain1992XPShandbook} In \Cref{fig:xps_f}, we report the F1s spectra with two subpeaks at 685.5 eV and 687.2 eV corresponding to LiF and F-C bonds, respectively. \cite{chastain1992XPShandbook, 2016_XPS_LN_ratios} In \Cref{fig:xps_mg}, we report the Nb4s, Li1s, and Mg2p spectra at 61.0 eV, 55.7 eV, and 50.7 eV, respectively (values are for bulk LN). \cite{chastain1992XPShandbook, 2016_XPS_LN_ratios} The Li1s peak energy agrees well with reported binding energies for LiF ($55.7 \pm 0.5$ eV). \cite{chastain1992XPShandbook, database2000nist} After ALE, we observe a 
0.3 eV shift for the Nb4s and Li1s peaks, and a 0.6 eV shift for the Mg2p peaks towards higher binding energies, as expected if fluoride bond formation occurred. \cite{chastain1992XPShandbook} There is also an increased concentration of Mg after ALE, suggesting that MgF$_2$ is also formed.

In \Cref{fig:xps_histogram}, we report the atomic concentrations of Nb, Li, Mg, O, and F obtained from the XPS data at various stages of the process. The atomic concentrations are normalized by the estimated carbon content for each sample, which is about 55\% and is due to presence of the conductive carbon coating. The uncertainties for all atomic concentration numbers (including C) were estimated from the CasaXPS software using a Monte Carlo routine, and the carbon-normalized uncertainties are propagated by adding uncertainties in quadrature. Surface lithium content is observed to increase after 1 H$_2$ plasma half-cycle. The presence of fluorine is likely from residual fluorine on the chamber walls after the chamber clean.  This trend differs from that reported in previous studies in which surface lithium concentration was found to decrease after a high power (500-2000 W), high temperature (170 \textdegree C), hour-long H$_2$ plasma exposure, forming proton-exchanged LN (c.f. Fig. 4 in Ref.~\cite{Ren_PE_LN_Plasma}). These plasma conditions in Ref.~\cite{Ren_PE_LN_Plasma} are more similar in terms of substrate temperature and exposure time to acid-based proton exchange, possibly accounting for the difference. After ALE, the fraction of F, Li, and Mg increases  compared to those of the untreated sample. Considering as well the peak energy shifts mentioned earlier, we hypothesize that LiF and MgF$_2$ are formed and redeposited on the surface owing to their low volatility.

\begin{table}
\caption{\label{tab:summary_table} Comparison of metrics from LN ALE recipes with different removal step plasma exposures on bulk LN. The bulk LN samples used have an initial surface roughness of $R_q = 0.2$ nm. Values are from a 40 second H$_2$ plasma exposure followed by a 40 second plasma exposure indicated in the table, over 20 cycles. Whether the O$_2$/SF$_6$ and Cl$_2$/BCl$_3$ processes were at saturation was not determined.}
\begin{ruledtabular}
\begin{tabular}{ccccc}
Plasma Type & EPC (nm/cycle) & RMS Roughness (nm) & Average Roughness (nm) & Synergy (\%) \\
\hline
SF$_6$/Ar   &  $1.59  \pm 0.02$              &    $ 0.57 \pm 0.18 $                  &        $ 0.44 \pm 0.11 $              &  96.9  \\
O$_2$/SF$_6$   &  $2.24 \pm 0.02$              &      $1.47 \pm 0.52$                &      $0.99 \pm 0.40$                &  99.5   \\
Cl$_2$/BCl$_3$  &  $1.65 \pm 0.03$              &      $0.90 \pm 0.46$                &     $0.59 \pm 0.20$               &  91.5    
\end{tabular}
\end{ruledtabular}
\end{table}

\begin{figure}
    \centering
    
{\includegraphics[width = 450pt]{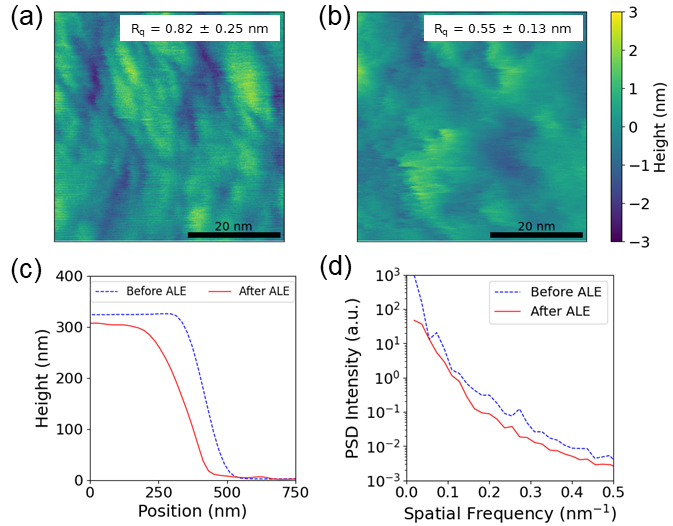}
    \phantomsubcaption\label{fig:before_ALE_scan}
    \phantomsubcaption\label{fig:after_ALE_scan}
    \phantomsubcaption\label{fig:wg_profile}
    \phantomsubcaption\label{fig:afm_psd}
    
}
\caption{AFM scan showing height-maps of TFLN waveguide sidewall with linear-plane tilt removal before (a) and after 50 ALE cycles (b). (c) Averaged AFM line scans of the TFLN waveguide side profile before and after 50 cycles ALE. The waveguide width decreases by 50 nm on each side, yielding a lateral etch rate of 1 nm/cycle, which is comparable to the vertical etch rate of 1.59 nm/cycle measured on bulk LN. (d) Height-map PSD of the samples before ALE and after 50 cycles of ALE.
}
\label{fig:afm} 
\end{figure}




We characterized the effect of all three ALE recipes on surface roughness of bulk LN samples. AFM scans over  500 $\times$ 500 nm$^2$ were obtained on bulk LN subjected to 20 cycles of ALE. Table~\ref{tab:summary_table} compares the EPC, surface roughness, and synergy of the three different recipes. For reference, the bulk LN samples  have an initial surface roughness of $R_q = 0.2$ nm. The O$_2$/SF$_6$ process yielded the roughest surface but the highest synergy, and the Cl$_2$/BCl$_3$ process yielded a similar etch rate to the SF$_6$/Ar process with the lowest synergy of the three recipes. It is hypothesized that the Cl$_2$/BCl$_3$ process may be modified to have higher synergy by lowering the RIE power on the Cl$_2$/BCl$_3$ plasma half step. In comparison, the SF$_6$/Ar process at saturation produced the smoothest surface after 20 cycles of ALE with $R_q$ of $0.57 \pm 0.18$ nm. The increase in surface roughness may be attributed to redeposition of LiF and MgF$_2$.

Since sidewall roughness of TFLN waveguides are rougher than the surface of bulk LN samples, we next characterized the effect of the SF$_6$/Ar ALE process on the sidewall surface roughness of TFLN waveguides. For these measurements, we used additional samples consisting of TFLN with Ar$^+$ milled waveguides that were smoothed post-etch with an HF dip and RCA clean, corresponding to the state-of-art process for TFLN device fabrication. \cite{Zhu:21, Ar_LN_Etch_Clean} \Cref{fig:before_ALE_scan,fig:after_ALE_scan} show the quadratic plane-fit height map of Ar$^+$ etched sidewall before and after 50 cycles of ALE, respectively. After ALE, the sidewall surface is visually smoother. The sidewall surface smoothing may be attributed to the isotropic nature of the etch. 

To support this hypothesis, we measured the lateral etch rate of the waveguide using AFM. \Cref{fig:wg_profile} shows an AFM profile  averaged over the whole scanned image of the TFLN waveguide sidewall before and after 50 cycles of ALE. From the decrease in width, we infer the lateral etch rate to be 1 nm/cycle on each side of the waveguide compared to a vertical etch rate of $1.59$ nm/cycle previously measured on bulk LN, confirming the largely isotropic nature of the process.




To more quantitatively characterize the surface topology of the TFLN sidewalls, we computed the surface power spectral density (PSD) from the AFM scans. \Cref{fig:afm_psd} shows the PSD curves before and after ALE on TFLN. Using AFM scans, the initial RMS sidewall roughness is measured to be $0.82 \pm 0.25$ nm and an average roughness of $0.65 \pm 0.16$ nm. After 50 cycles, $R_q$ and $R_a$ are measured as $0.55 \pm 0.13$ nm and $0.44 \pm 0.12$ nm, respectively. The PSD is observed to decrease over all measured spatial frequencies. Therefore, despite the roughening observed on flat LN surfaces, sidewall smoothing is still observed owing to the isotropic nature of the process.


\section{Discussion}

Our ALE process may find applications in improving the photonic performance of TFLN devices by reducing optical loss associated with corrugations in PPLN and sidewall roughness. The etch rate of a typical RCA wet etch exhibits uncontrolled variability due to temperature and concentration fluctuations in solution. The reported ALE process has potential to overcome these issues due to the self-limiting nature of the process with well-controlled etch rates. 


Future topics of interest include investigating the mechanism for etch selectivity of the hydrogen-exposed surface over the unmodified surface and identifying approaches to reduce the quantity of redeposited Li and Mg compounds. A post-ALE wet clean may be beneficial to remove the redeposited compounds selectively, in contrast to the present approach using an RCA wet etch which etches lithium niobate. Development of an in-situ gas-based removal process or a process based on thermal cycling may enable redeposition-free ALE. For thermally cyclic processing, investigation of chemistries which produce more volatile products, such as those based on Br, is of interest for further study. Directional ALE processes with high anisotropy are also of interest as they  could be employed for pattern transfer, yielding precise and uniform control of etch depth over the entire chip with precision of around the EPC ($\sim$1 nm). This degree of control would permit scaling of TFLN devices and circuits to the system level. The ALE system in our work (Oxford Instruments, Plasma Pro 100 Cobra) is able to process 150 mm diameter substrates, and therefore our process has the potential to extend to wafer-scale applications. 




\section{Conclusion}
We have reported an isotropic ALE process for $x$-cut MgO-doped lithium niobate consisting of sequential exposures of hydrogen plasma and SF$_6$/Ar plasma that is compatible with low-pressure ICP RIE systems. We observe an etch rate of $1.59 \pm 0.02$ \AA/cycle with a synergy exceeding $96$\%. Both half-steps exhibited saturation with respect to exposure time, though the SF$_6$ plasma half step was observed to soft-saturate. The substitution of O$_2$/SF$_6$ or Cl$_2$/BCl$_3$ plasmas in place of SF$_6$/Ar plasma was also found to yield ALE with synergies exceeding 90\%. Finally, the process was found to smoothen the sidewalls of TFLN waveguides fabricated using the state-of-art process, suggesting the potential of ALE to enhance the photonic performance of TFLN devices.

\section{Acknowledgements}
This work was supported by Oxford Instruments and the NSF under Award \#2234390. This research was carried out, in part, at the Jet Propulsion Laboratory (JPL), California Institute of Technology, under contract with the National Aeronautics and Space Administration (NASA). We gratefully acknowledge the critical support and infrastructure provided for this work by The Kavli Nanoscience Institute and the Molecular Materials Research Center of the Beckman Institute at the California Institute of Technology. 

\section{Author Declarations}
\subsection{Conflict of Interest}
I.I.C., J.S., R.S., A.M., F.G., and A.J.M. have submitted a United States provisional patent on this technology.

\section{Data Availability}
The data that support the findings of this study are available from the corresponding author upon reasonable request.

\section*{References}

\bibliographystyle{apsrev4-2}

\bibliography{refs}

\end{document}